\documentclass[doublecol]{epl2}
\usepackage{bm}
\usepackage{amsmath}
\usepackage{amssymb}

\usepackage{color}
\usepackage{ulem}

\newcommand{\INS}[1]{{{#1}}}
\newcommand{\KILL}[1]{}

\newcommand{\COM}[1]{}

\title{Design and control of noise-induced synchronization patterns}
\shorttitle{Design and control of noise-induced synchronization patterns}
\author{W. Kurebayashi\inst{1}\hspace{-0.2cm}\footnotemark[1] \and T. Ishii\inst{1} \and M. Hasegawa\inst{2} \and H. Nakao\inst{1}}
\shortauthor{W. Kurebayashi \etal}

\institute{                    
  \inst{1} Graduate School of Information Science and Engineering,
	Tokyo Institute of Technology - 2-12-1 O-okayama, Meguro-ku, Tokyo 152-8552, Japan\\
  \inst{2} Department of Electrical Engineering, Tokyo University of Science -
	6-3-1 Niijuku, Katsushika-ku, Tokyo 125-8585, Japan
}
\pacs{nn.mm.xx}{First pacs description}
\pacs{nn.mm.xx}{Second pacs description}
\pacs{nn.mm.xx}{Third pacs description}

\abstract{
We propose a method for controlling synchronization patterns of limit-cycle oscillators by common noisy inputs,
i.e.,~by utilizing noise-induced synchronization.
Various synchronization patterns, including fully synchronized and clustered states,
can be realized by using linear filters that generate appropriate common noisy signals from given noise.
The optimal linear filter can be determined from the linear phase response property of the oscillators and the power spectrum of the given noise.
The validity of the proposed method is confirmed by numerical simulations.
}

\begin{document}

\maketitle

\footnotetext[1]{kurebayashi.w.aa@m.titech.ac.jp}

\section{Introduction}

Various nonlinear dynamical systems tend to synchronize when driven by a common noisy input.
This phenomenon, called noise-induced synchronization, is observed in many systems, for example,
in neurons~\cite{mainen95__reliab_of_spike_timin_in_neocor_neuron,neiman2002,
galan06__correl_induc_synch_of_oscil,galan2008},
electric circuits~\cite{yoshida06__noise_induc_synch_of_uncoup_nonlin_system},
electronic devices~\cite{utagawa08__noise_induc_synch_among_sub},
microbial cells~\cite{zhou05__molec_commun_throug_stoch_synch},
lasers~\cite{uchida04__consis_of_nonlin_system_respon},
and chaotic dynamical systems~\cite{toral2001,zhou2003}.
{It has been clarified that noise-induced synchronization has quite a different mechanism from phase locking to periodic forcing,
i.e., the oscillators are not entrained by the input but still exhibit {mutual} synchronization, characterized by coherent distributions of the phase differences.}
Analytical investigations of this phenomenon for limit-cycle oscillators can be performed by using the phase reduction method~\cite{kuramoto1984},
and its properties have been widely studied in the last decade~\cite{teramae04__robus_of_noise_induc_phase,
Goldobin05a,Goldobin05b,
nakao05__synch_of_limit_cycle_oscil,
nakao07__noise_induc_synch_and_clust,yoshimura08__invar_of_frequen_differ_in,marella2008,hata10__synch_of_uncoup_oscil_by,%nagai10__noise_induc_synch_of_large,
goldobin2010,kure2012,burton2012,ming2013,zhou2013}.

\INS{In contrast to phase locking that requires periodic forcing whose frequency is close to rational multiples of the natural frequency of the oscillators,
noise-induced synchronization can occur even for white noise~\cite{teramae04__robus_of_noise_induc_phase,
Goldobin05a,Goldobin05b,
nakao05__synch_of_limit_cycle_oscil,
nakao07__noise_induc_synch_and_clust,yoshimura08__invar_of_frequen_differ_in,marella2008,%nagai10__noise_induc_synch_of_large,
goldobin2010,kure2012,burton2012,ming2013,zhou2013}.
Thus, noise-induced synchronization may more easily be realized than phase locking, because common environmental noise is ubiquitous in nature.
%%
%% while periodic forcing with appropriate frequency may not always exist.
%%
Indeed, it is conjectured that some plants utilize common environmental noise for synchronization to realize biological functions~\cite{satake2002, lyles2009}.
%If so, some biological mechanism for filtering the environmental noise may exist to improve mutual synchronization.
In such biological systems, some kind of filtering mechanisms for the environmental noise may also exist to improve noise-induced synchronization.
Such filtering mechanisms, if any, would also be useful in practical applications such as noise-induced synchronization of sensor networks~\cite{yasuda2013}.
The aim of this letter is to provide a theoretical basis for optimizing noise-induced synchronization by filtering the given noise.
}

In our previous works~\cite{nakao07__noise_induc_synch_and_clust,kure2012},
we developed quantitative theories that predict global statistical properties of the noise-induced synchronization,
such as formation of various synchronization patterns and fluctuations around these patterns,
not only whether the oscillators synchronize with the others.
{The synchronization patterns are characterized by the probability density function (PDF) of the phase differences between the oscillators,
{which can be} calculated from the phase response property of the oscillators and the statistical properties of the noisy inputs.}

In this letter, on the basis of ref.~\cite{kure2012}, we propose a method for designing and controlling
various noise-induced synchronization patterns of limit-cycle oscillators, including the fully synchronized and clustered states.
%%
%We can design the synchronization pattern by optimizing the noisy input that drives the oscillators,
%because the statistical properties of the noisy input determine the synchronization pattern of the driven oscillators, i.e., the distribution of their phase difference.
%%
{Since the synchronization pattern of the oscillators
depends on {the} statistical properties of the noisy input, we can design the synchronization pattern by optimizing the noisy input
so that some {objective function}, e.g., the degree of synchronization, is maximized.}
We develop an optimization method for the noisy input, namely, for a linear filter that transforms given noise into an appropriate noisy input so that the desired synchronization pattern is realized.
The validity of the proposed method is confirmed by numerical simulations.

\section{Model}

We consider an ensemble of $N$ uncoupled identical limit-cycle oscillators
subjected to correlated noise and independent noise,
described by the following Langevin equations:
\begin{align}
\dot{\bm{X}}_j(t) &= \bm{F}(\bm{X}_j) + \epsilon \bm{G}(\bm{X}_j) [ I_{j}(t) + \zeta_j(t) ],  
 \label{eq. lco}
\end{align}
for $j=1,\ldots,N$. Here, $\bm{X}_j(t) \in\mathbb{R}^{n}$ is the state of the oscillator $j$ at time $t$,
$\bm{F}(\bm{X}_j) \in\mathbb{R}^{n}$ is a vector field representing the oscillator dynamics,
$\bm{G}(\bm{X}_j) \in\mathbb{R}^{n}$ represents the coupling of the oscillator to the noisy inputs,
$I_{j}(t) \in \mathbb{R}$ is the correlated noise, $\zeta_{j}(t) \in \mathbb{R}$ is the independent noise that drives each oscillator independently, and
$\epsilon$ ($\ll1$) is a small parameter that controls the strength of the noisy inputs.
We assume that eq.~(\ref{eq. lco}) possesses a stable limit-cycle orbit $\bm{X}_0(t)$ with period $T$ and frequency $\omega:=2\pi/T$ when $\epsilon=0$.
 The correlated noise is generated from two given noisy signals by a linear filter as $I_{j}(t) = f * (\xi(t) + \eta_{j}(t))$,
where the noise $\xi(t)\in \mathbb{R}$ is common to all oscillators, $\eta_{j}(t)\in \mathbb{R}$ is independently applied to each oscillator,
$f(\tau) \in\mathbb{R}$ is a  filter function that
transforms the given noise to appropriate noise for realizing desired synchronization patterns,
and the star ($*$) denotes convolution $f * \alpha(t) = \int_{-\infty}^{+\infty} f(\tau) \alpha(t - \tau) d\tau$.
We introduced two independent noise terms $\eta_{j}(t)$ and $\zeta_{j}(t)$ to take into account the effect of external disturbances before and after filtering.

For example, the oscillators described by eq.~(\ref{eq. lco}) can be regarded as spiking neurons receiving artificial injection currents as in ref.~\cite{galan06__correl_induc_synch_of_oscil}.
In this case, the filtered noise $I_{j}(t)$ represents the injected current to each neuron, and we may suppose $\eta_j(t)=0$.
The filter $f(\tau)$ is used for generating an appropriate injection current,
and $\zeta_j(t)$ is independent noise inherent in each neuron, e.g.,~channel or synaptic noise.
In the situation of ref.~\cite{yasuda2013} where noise-induced synchronization of wireless sensor networks is considered, each oscillator described by eq.~(\ref{eq. lco}) corresponds to each sensor node.
The sensor node measures a noisy environmental signal $\xi(t)+\eta_j(t)$,
and the filter $f(\tau)$ implemented on each sensor node transforms the signal into an appropriate noisy input $I_{j}(t)$ {that induces} synchronization of the sensor nodes.
We may suppose $\zeta_j(t)=0$ in this case.

We assume that $\xi(t)$, $\eta_j(t)$, and $\zeta_j(t)$ are mutually independent zero-mean Gaussian noise,
i.e.,~$\langle \xi(t) \rangle = \langle \eta_{j}(t) \rangle = \langle \zeta_{j}(t) \rangle = 0$ and $\langle\xi(t)\eta_j(t-\tau)\rangle = \langle\xi(t)\zeta_j(t-\tau)\rangle
= \langle\eta_j(t)\eta_k(t-\tau)\rangle = \langle\zeta_j(t)\zeta_k(t-\tau)\rangle
= \langle\eta_k(t)\zeta_\ell(t-\tau)\rangle  = 0$ for any $j$, $k$, and $\ell$ ($j \neq k$),
where $\langle\cdot\rangle$ denotes the ensemble average over realizations of $\xi(t)$, $\eta_j(t)$, and $\zeta_j(t)$.
For simplicity, we assume that the statistical properties of $\eta_j(t)$ and $\zeta_j(t)$ do not depend on the oscillator index $j$.
Their power spectra are given by
$P_{\xi}(\Omega) := \int_{-\infty}^{+\infty} e^{-i\Omega\tau} \langle\xi(t)\xi(t-\tau)\rangle d\tau$,
$P_{\eta}(\Omega) := \int_{-\infty}^{+\infty} e^{-i\Omega\tau} \langle\eta_j(t)\eta_j(t-\tau)\rangle d\tau$,
and $P_{\zeta}(\Omega) := \int_{-\infty}^{+\infty} e^{-i\Omega\tau} \langle\zeta_j(t)\zeta_j(t-\tau)\rangle d\tau$.
We also define the amplitude response of the filter $f(\tau)$ as $A(\Omega) := |\int_{-\infty}^{+\infty} e^{-i\Omega\tau} f(\tau) d\tau|$.

By the phase reduction method~\cite{kuramoto1984}, we can reduce the high-dimensional oscillator dynamics described by eq.~(\ref{eq. lco})
to a one-dimensional phase equation for small $\epsilon$,
\begin{align}
 \dot{\theta}_j &= \omega + \epsilon Z(\theta_j) [ f * (\xi_j(t) + \eta_j(t)) + \zeta_j(t) ] \cr
 &\quad + \epsilon^2 \nu(\theta_j) + O(\epsilon^3),
 \label{eq. pe}
\end{align}
where $\theta_j(t) \in [0,2\pi)$ is the phase of the oscillator $j$,
$Z(\theta_{j})$ is a sensitivity function that characterizes the response of the oscillator phase to noisy inputs,
and $\nu(\theta_j)$ represents the effect of amplitude relaxation dynamics of stochastic limit-cycle oscillators~\cite{yoshimura2008,goldobin2010}
(this term eventually vanishes and does not play a role in the following argument).
The sensitivity function $Z(\theta)$ is given as $Z(\theta) = \bm{G}^{\top}(\bm{X}) \nabla_{\bm{X}}\theta(\bm{X})|_{\bm{X}=\bm{X}_0(\theta/\omega)}$,
where $\theta(\bm{X})$ is the isochron of the limit cycle and $\nabla_{\bm{X}}\theta(\bm{X})|_{\bm{X}=\bm{X}_0(\theta/\omega)}$ represents its gradient
on the limit-cycle orbit at phase $\theta$~\cite{kuramoto1984}.

\section{Characterization of synchronization patterns}

As discussed in refs.~\cite{nakao07__noise_induc_synch_and_clust,kure2012},
the phase difference between two oscillators $\phi_{j,k}:=\theta_j-\theta_k$ characterizes the noise-induced synchronized state.
Since statistical properties of $\eta_j(t)$ and $\zeta_j(t)$ do not depend on the oscillator index $j$,
the PDF of the phase difference $\phi_{j,k}$ does not depend on the indices $j$ and $k$.
Thus, in the following, we denote the phase difference by $\phi$ without the oscillator indices.

{In our previous work~\cite{kure2012}, we obtained the stationary PDF $U(\phi)$ of the phase difference $\phi$
by employing effective white-noise approximation of the phase equations (\ref{eq. pe}) subjected to correlated colored noise and 
by deriving an averaged Fokker-Planck equation for $\phi$ from the multivariate Fokker-Planck equation for the phase variables $\{\theta_{j}\}$. 
It turns out that the correlation functions of the noise play an important role, and $U(\phi)$ is explicitly given by}
\begin{align}
 U(\phi) &= \frac{1}{\bar{u}} \cdot \frac{1}{g(0)-g(\phi)+h(0)},
 \label{eq. dist}
\end{align}
where $\bar{u} \in \mathbb{R}$ is a normalization constant determined by $\int_{-\pi}^{+\pi}U(\phi)d\phi=1$, and $g(\phi) \in \mathbb{R}$ and $h(\phi) \in \mathbb{R}$ are correlation functions
{of the noise terms in eq.~(\ref{eq. pe}), defined as
$g(\phi)=\int_{-\infty}^{+\infty}\langle Z(\theta(t))f*\xi(t) Z(\theta(t-\tau)+\phi)f*\xi(t-\tau) \rangle d\tau$
and
$h(\phi)=\int_{-\infty}^{+\infty}\langle Z(\theta(t))[f*\eta_j(t)+\zeta_j(t)] Z(\theta(t-\tau)+\phi)[f*\eta_j(t)+\zeta_j(t-\tau)] \rangle d\tau$.
}In Fourier representation, these functions can be written as
$g(\phi) = \sum_{\ell=-\infty}^{+\infty} g_{\ell} e^{i\ell\phi}$
and
$h(\phi) = \sum_{\ell=-\infty}^{+\infty} h_{\ell} e^{i\ell\phi}$,
where the Fourier coefficients $g_{\ell}, h_{\ell} \in \mathbb{R}$ are given by
\begin{align}
	g_{\ell} &= |z_{\ell}|^2 |A(\ell\omega)|^2 P_{\xi}(\ell\omega), \quad \cr
	h_{\ell} &= |z_{\ell}|^2 |A(\ell\omega)|^2 P_{\eta}(\ell\omega) + |z_{\ell}|^2 P_{\zeta}(\ell\omega),
 \label{eq. corr func}
\end{align}
and $z_{\ell} := \frac{1}{2\pi} \int_{-\pi}^{+\pi} e^{-i\ell\theta} Z(\theta) d\theta$
is the Fourier coefficient of $Z(\theta)$.
From eq.~(\ref{eq. dist}), we see that the PDF $U(\phi)$ is symmetric about $\phi=0$ and has a maximum at $\phi=0$.
For example, a PDF $U(\phi)$ with a single peak at $\phi=0$ represents the synchronized state of the oscillators,
and $U(\phi)$ with $k$ peaks represents the $k$-clustered state.
{Thus, when the amplitude response $A(\Omega)$ of the filter $f(\tau)$ enhances the $k$-th mode of the correlation function $g(\phi)$ in eq.~(\ref{eq. corr func}),  $k$-clustered distribution is emphasized in the PDF $U(\phi)$.}

\section{Design of synchronization patterns}

Equation~(\ref{eq. dist}) indicates that we can design the stationary PDF of the phase difference $U(\phi)$,
i.e., the synchronization pattern, by varying the correlation function $g(\phi)$.
Therefore, given the power spectra $P_{\xi}(\Omega)$, $P_{\eta}(\Omega)$, $P_{\zeta}(\Omega)$, and the sensitivity function $Z(\theta)$,
we can try to find an optimal filter $f(\tau)$ that gives $g(\phi)$.
%%
%To design $f(\tau)$, we introduce an objective function $R\{f\}$ that evaluates the fitness of the filter $f(\tau)$
%for realizing the desired synchronization pattern as
%%
{In this study, rather than explicitly specifying the precise PDF of the oscillators as the target, we aim to maximize its statistical property,
e.g., the degree of synchronization or clustering.
This is because such macroscopic properties, rather than precise functional forms of $U(\phi)$, are relevant in practical applications. 
Note also that we cannot generate arbitrary PDFs {but} only optimize the PDF given in the form of eq.~(\ref{eq. dist}).}

{To design the optimal filter $f(\tau)$, we introduce the following objective functional $R\{f\}$
of $f(\tau)$ characterizing the statistical property of $U(\phi)$ as a measure for choosing an optimal
synchronization pattern:}
\begin{align}
 R\{f\} &= \int_{-\pi}^{+\pi} U(\phi;f) q(\phi) d\phi,
 \label{eq. R}
\end{align}
{where we explicitly show the dependence of $U(\phi;f)$ on $f$.  The function $q(\phi)$ determines what statistical property we focus on.}
{We try to design synchronization patterns with desired statistical properties by choosing appropriate $q(\phi)$.}

{In the numerical simulations given below, we will use the following functions for designing the synchronization patterns:
$q_1(\phi) = \cos\phi, \ q_2(\phi) = \delta(\phi), \ q_3(\phi) = \cos3\phi$, and $q_{4}(\phi) = \cos2\phi$,
where $\delta(\cdot)$ denotes the Dirac delta function.
When we use $q_1(\phi)$, the objective functional $R\{f\}$ corresponds to the order parameter introduced in ref.~\cite{marella2008},
which characterizes the degree of noise-induced synchronization.
When we use $q_2(\phi)$, the objective functional $R\{f\}$ corresponds to the maximum of the PDF $U(\phi)$ at $\phi=0$,
which also characterizes the degree of synchronization, but in a more strict way, i.e., it counts only the oscillator pairs with exactly zero phase difference.
When we use $q_3(\phi)$, the objective functional $R\{f\}$ characterizes three-clustered states,
in which three synchronized subgroups of oscillators are formed.}
{Similarly, $q_4(\phi)$ {characterizes} two-clustered states.}

In the following, we represent $q(\phi)$ as a Fourier series, $q(\phi) = \tilde{q}_0 + 2\sum_{\ell=1}^{\infty} \tilde{q}_\ell \cos \ell\phi,$
where the coefficient $\tilde{q}_\ell \in \mathbb{R}$ represents the weight of the $\ell$-th Fourier mode.
Expanding $U(\phi)$ as $U(\phi)=\frac{1}{2\pi} + \sum_{\ell=1}^{\infty} \tilde{u}_\ell \cos \ell\phi$, eq.~(\ref{eq. R}) can be written as
$R\{f\} = \sum_{\ell=1}^{\infty} \tilde{q}_\ell\tilde{u}_\ell.$
By finding optimal $\tilde{u}_{\ell}$ for given $\tilde{q}_{\ell}$, we can obtain a PDF $U(\phi)$ and a filter $f(\tau)$ that maximizes the objective functional $R\{f\}$.

\section{Optimization of the filter}

By maximizing the objective functional $R\{f\}$, we seek for the optimal filter $f(\tau)$.
However, unconstrained maximization of $R\{f\}$ often leads to divergent $f(\tau)$.
We also need to take into account that our present theory is not valid for strong noisy inputs,
because the phase reduction method requires the input given to the oscillators to be sufficiently weak~\cite{kuramoto1984}.
Thus, we should introduce some constraint on the filter function $f(\tau)$.

In this study, we formulate the constrained optimization problem of the objective functional $R\{f\}$ as follows:
\begin{align}
 &\underset{f}{\rm maximize} && R\{f\},
 \label{eq. opt1} \\
 &{\rm subject\ to} && \sigma^2 := \langle I_{j}(t)^2 \rangle 
  = C,
 \label{eq. opt2}
\end{align}
where the condition eq.~(\ref{eq. opt2}) constrains the variance $\sigma^{2}$ of the filtered noise $I_{j}(t)$ to be a constant $C$.
Using the power spectra $P_{\xi}(\Omega)$ and $P_{\eta}(\Omega)$ and the amplitude response $A(\Omega)$ of the filter $f(\tau)$,
the variance $\sigma^2$ can be written as
\begin{align}
 \sigma^2 = \int_{-\infty}^{+\infty} |A(\Omega)|^2 [P_{\xi}(\Omega) + P_{\eta}(\Omega)] d\Omega.
 \label{eq. sigma3}
\end{align}
By solving the optimization problem described by eqs.~(\ref{eq. opt1}) and (\ref{eq. opt2}), we can, in principle, obtain the optimal filter $f(\tau)$
for maximizing the objective functional $R\{f\}$.

Actually, we should also take into account that the optimal solution of eqs.~(\ref{eq. opt1}) and (\ref{eq. opt2}) may not be implemented in practice.
The optimal amplitude response $A(\Omega)$ obtained as above often has delta peaks at $\Omega=\ell\omega$ ($\ell \in \mathbb{Z}$),
{i.e., $|A(\Omega)|^2 = a_0 \delta(\Omega) + \sum_{\ell=1}^{\infty} a_\ell [\delta(\Omega-\ell\omega) + \delta(\Omega+\ell\omega)]$
($a_\ell$ is some coefficient and $\omega$ is the natural frequency of the oscillator),}
because the PDF $U(\phi)$ depends only on the harmonic components $P_{\xi}(0),P_{\xi}(\omega),P_{\xi}(2\omega),\ldots$
and $P_{\eta}(0),P_{\eta}(\omega),P_{\eta}(2\omega),\ldots$ of the noise (see eqs.~(\ref{eq. dist}) and~(\ref{eq. corr func})).
{Such a delta-peaked amplitude response $A(\Omega)$ corresponds to a physically unrealistic filter that extracts only purely harmonic components from the noise,
which leads to phase locking rather than noise-induced synchronization of the oscillators.
Besides, such {singular} $A(\Omega)$ cannot be realized in practical implementation of the linear filter $f(\tau)$.}

To overcome this problem, we restrict the class of $A(\Omega)$ and further assume that the square of the amplitude response is expressed
{as a finite sum of narrow-band basis functions} as
\begin{align}
 |\tilde{A}(\Omega)|^2 &:= \sum_{\ell=-m}^{m} c_{|\ell|} W(\Omega - \ell\omega),
 \label{eq. sigma2}
\end{align}
where $c_\ell$ ($\ell=0, 1, 2, \ldots$) is a weight coefficient,
$W(\Omega)$ represents {a narrow-band basis function prespecified before the optimization process, e.g., a Gaussian function,}
and $m$ is the maximum wavenumber of the filter. 
{We assume that the basis function $W(\Omega)$ is localized in the range $|\Omega| < \omega$, i.e., $W(\Omega) \approx 0$ holds for $|\Omega|\ge\omega$.
The parameter $m$ should be sufficiently large to obtain a good filter.
The restricted amplitude response $\tilde{A}(\Omega)$ {is experimentally feasible},
because $W(\Omega - \ell\omega)$ in eq.~(\ref{eq. sigma2}) can be implemented by a band-pass filter that passes frequencies around $\Omega = \ell\omega$.}

We introduce a new parameter $\bm{\beta}=[\beta_0,$ $\beta_1,\ldots,$ $\beta_{m}]^{\top}\in\mathbb{R}^{m+1}$
as $\beta_0=\sqrt{|b_0 c_0|}$ and $\beta_\ell=\sqrt{2|b_\ell c_\ell|}$ for $\ell=1,\ldots,m$,
where $b_\ell$ ($\ell=0,1,2,\ldots$) is defined as $b_\ell=\int_{-\infty}^{+\infty}W(\Omega-\ell\omega)[P_{\xi}(\Omega) + P_{\eta}(\Omega)]d\Omega$.
 The optimization problem~(\ref{eq. opt1}) and (\ref{eq. opt2}) with the above restriction can then be expressed as
\begin{align}
 &\underset{\bm{\beta}}{\rm maximize} && \tilde{R}(\bm{\beta}) := \int_{-\pi}^{+\pi}\tilde{U}(\phi;\bm{\beta})q(\phi)d\phi,
 \label{eq. opt2-1} \\
 &{\rm subject\ to} && \sigma^2 \approx ||\bm{\beta}||^2 = \sum_{l=0}^{m-1} |\beta_\ell|^2 = C,
 \label{eq. opt2-2}
\end{align}
where $\tilde{U}(\phi;\bm{\beta})$ is the PDF of the phase difference $\phi$ obtained {by plugging eq.~(\ref{eq. sigma2}) into eq.~(\ref{eq. corr func})},
and eq.~(\ref{eq. opt2-2}) follows from eqs.~(\ref{eq. sigma3}) and (\ref{eq. sigma2}) and the definition of $\bm{\beta}$.
Thus, we can employ $\bm{\beta}$ as a design parameter of the optimization problem described by eqs.~(\ref{eq. opt2-1}) and (\ref{eq. opt2-2}).

\section{Optimization algorithm}

To solve the optimization problem given by eqs.~(\ref{eq. opt2-1}) and (\ref{eq. opt2-2}), we use the gradient descent algorithm.
We randomly choose an initial value $\bm{\beta}^{(0)}$ and iteratively calculate
$\bm{\beta}^{(j)}$ for $j \ge 1$, where $\bm{\beta}^{(j)}$ is the design parameter $\bm{\beta}$  at  $j$-th iteration.
At each iteration, we update $\bm{\beta}^{(j)}$ as
\begin{align}
 \tilde{\bm{\beta}}^{(j+1)} &= \bm{\beta}^{(j)} + \alpha \nabla_{\bm{\beta}}\tilde{R}(\bm{\beta}^{(j)})
 \label{eq. fpa}
\end{align}
and normalize it as
$\bm{\beta}^{(j+1)} = \sqrt{C}\tilde{\bm{\beta}}^{(j+1)}/||\tilde{\bm{\beta}}^{(j+1)}||,$
so that $\bm{\beta}^{(j)}$ satisfies the constraint~(\ref{eq. opt2-2}). 
Here, $\alpha$ is a constant that controls the step size,
and $\nabla_{\bm{\beta}}\tilde{R}(\bm{\beta}) \in\mathbb{R}^{m+1}$ represents the gradient of $\tilde{R}(\bm{\beta})$ with respect to $\bm{\beta}$,
i.e.,~$\nabla_{\bm{\beta}}${$\tilde{R}(\bm{\beta}) = [\frac{\partial \tilde{R}(\bm{\beta})}{\partial \beta_0},$ $\ldots,$ $\frac{\partial \tilde{R}(\bm{\beta})}{\partial \beta_{m}}]^{\top}$.

For simplicity of notation, we define $u(\phi) := 1/[g(0) - g(\phi) + h(0)]$.
{The normalization constant is given by $\bar{u}=\int_{-\pi}^{+\pi}u(\phi)d\phi$.}
Then, the gradient $\frac{\partial \tilde{R}(\bm{\beta}^{(j)})}{\partial \beta_\ell}$ in eq.~(\ref{eq. fpa}) can be expressed as
$\frac{\partial \tilde{R}(\bm{\beta})}{\partial \beta_\ell}
= \int_{-\pi}^{+\pi} \frac{1}{\bar{u}^2} [\frac{\partial u(\phi)}{\partial \beta_\ell} \bar{u}
- u(\phi) \frac{\partial\bar{u}}{\partial \beta_\ell} ] q(\phi) d\phi,$
and the gradients $(\partial u(\phi)/\partial \beta_\ell)$ and $(\partial\bar{u}/\partial \beta_\ell)$ can be calculated
from eqs.~(\ref{eq. dist}), (\ref{eq. corr func}) and (\ref{eq. sigma2}) and the definition of $\bm{\beta}$ as 
$\frac{\partial u(\phi)}{\partial \beta_\ell}
 = 2 u^2(\phi) \sum_{k=0}^{m-1} W(k\omega - \ell\omega)
\frac{\beta_\ell|z_k|^2}{b_\ell} [ P_\xi(k\omega)(\cos k\phi-1) - P_\eta(k\omega) ],$
$\frac{\partial\bar{u}}{\partial \beta_\ell}
= \int_{-\pi}^{+\pi} \frac{\partial u(\phi)}{\partial \beta_\ell} d\phi.$

Because $W(\Omega)$ is localized in the range $|\Omega| < \omega$, the above expression can be simplified as follows:
$\frac{\partial u(\phi)}{\partial \beta_\ell}
= 2 u^2(\phi) W(0) \frac{\beta_\ell|z_\ell|^2}{b_\ell} \left[ P_\xi(\ell\omega)(\cos \ell\phi-1) - P_\eta(\ell\omega) \right],$
which reduces the computational cost of the optimization process.
Using the optimized $\bm{\beta}$, we can obtain the optimal amplitude response $\tilde{A}(\Omega)$
as the square root of eq.~(\ref{eq. sigma2}), whose coefficients $c_0, \ldots, c_{m}$ are given by
$c_0 = \frac{|\beta_0|^2}{b_0},\ c_\ell = \frac{|\beta_\ell|^2}{2b_\ell},$ for $\ell=1,2,\ldots,m$.
Thus, using, e.g.,~the least-squares method~\cite{filter_design},
we can calculate the  optimal filter $f(\tau)$ from the amplitude response $\tilde{A}(\Omega)$.

\section{Numerical simulations}

To confirm the validity of our method, we performed numerical simulations using several examples of the objective functionals.
In the first example, we use the FitzHugh-Nagumo (FHN) model of a periodically firing neuron.
This model has a two-dimensional state variable
$(u, v)$, which obeys $\dot{v}(t) = v - v^3/3 - u + I_0 + I(t)$ and $\dot{u}(t) = \mu (v + c - du)$, $\mu=0.08$, $c=0.7$, $d=0.8$, and $I_{0}=0.875$.
The frequency of the oscillation is approximately $\omega = 0.173$ and the noisy input $I_j(t)$ is given to $v(t)$.
The sensitivity function $Z(\theta)$ to $I(t)$ and its Fourier coefficients are shown in figs.~\ref{fig.1} (a) and (b).

As the noisy inputs, we use the Ornstein-Uhlenbeck noise, whose power spectra are given by
$P_{\xi}(\Omega)=\tilde{P}_{\rm OU}(\Omega;0.5)$, $P_{\eta}(\Omega)=P_{\zeta}(\Omega)=0.1\cdot\tilde{P}_{\rm OU}(\Omega;0.5)$,
with $\tilde{P}_{\rm OU}(\Omega;\gamma):=\gamma^2/(\gamma^2+\Omega^2)$ [fig.~\ref{fig.1} (c)].
As the basis function $W(\Omega)$, we employ a rectangular function,
$W(\Omega) = 1\ (|\Omega|<\omega/2), \ 0\ ({\rm otherwise}).$
The other parameters are set as follows: the coupling strength to the noise is $\epsilon=0.01$, the variance of the filtered noise is $C=10$, 
and the maximum wavenumber of the filter is $m=5$
(because $|z_{\ell}|$ is small when $\ell \geq 6$).
The parameter $\alpha$ used for the gradient descent is $\alpha=0.5$.
Note that the gradient descent algorithm finds only a local optimum and does not guarantee global optimality.
In order to obtain the global optimum, the algorithm should be repeated from sufficiently many initial states $\bm{\beta}^{(0)}$.

Figure~\ref{fig.2} shows the numerical results for the FHN model.
For the functions $q_1(\phi)$ and $q_2(\phi)$ defined previously,  synchronized states are successfully formed [figs.~\ref{fig.2} (b) and (d)].
For the sinusoidal $q_1(\phi)$, we obtain a filter that emphasizes only the first Fourier mode [fig.~\ref{fig.2} (a)] (see eq.~(\ref{eq. corr func})),
which results in a bell-shaped PDF $U(\phi)$ [fig.~\ref{fig.2} (b)].
For the delta-shaped $q_2(\phi)$, in contrast, we obtain a nontrivial filter that consists of multiple modes [fig.~\ref{fig.2} (c)].
In this case, the PDF $U(\phi)$ has a sharper peak than that for $q_1(\phi)$ [fig.~\ref{fig.2} (d)] and a more precisely synchronized state is realized.
Note that the high-frequency components of $\tilde{A}(\Omega)$ in fig.~\ref{fig.2} (c), which are stronger than the low-frequency components,
do not significantly affect the statistical property of $I_j(t)$, because $P(\ell\omega)$ is sufficiently small for large $\ell$.
Therefore, we can safely neglect the high-frequency components of $\tilde{A}(\Omega)$ whose wavenumbers are larger than $m$.
For the function $q_3(\phi)$, we obtain a filter that emphasizes only the third Fourier mode [fig.~\ref{fig.2} (e)],
which yields a three-clustered state as expected [fig.~\ref{fig.2} (f)].
Note that we cannot form a two-clustered state in the FHN model,
because {the phase response property of this model has odd symmetry, i.e.,}
the second Fourier coefficient $|z_2|$ of $Z(\theta)$  is {vanishingly} small as shown in fig.~\ref{fig.1} (b).

\begin{figure}
\onefigure[width=8cm]{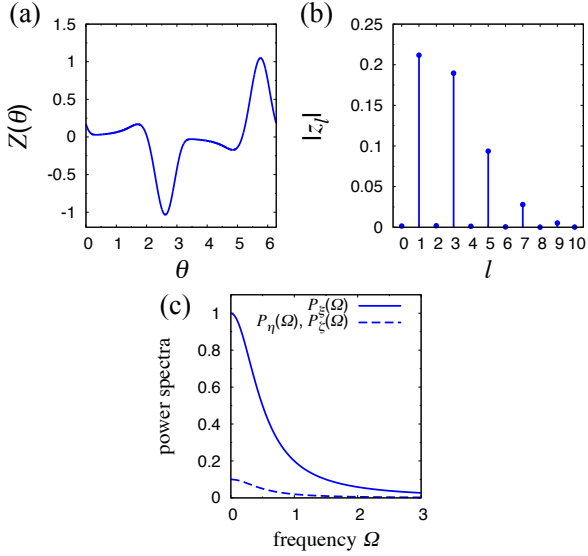}
 \caption{FitzHugh-Nagumo model.
 (a) Sensitivity function  $Z(\theta)$,
 (b) Fourier coefficients $|z_\ell|$, and
 (c) power spectra of the noisy inputs $P_{\xi}(\Omega)$, $P_{\eta}(\Omega)$ and $P_{\zeta}(\Omega)$.}
 \label{fig.1}
\end{figure}

\begin{figure}
\onefigure[width=8cm]{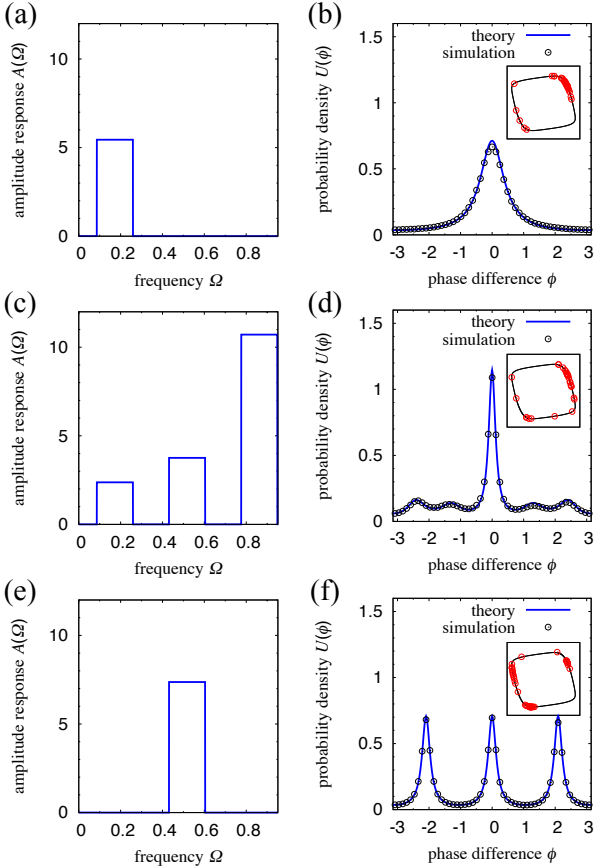}
 \caption{FitzHugh-Nagumo model.
 [(a), (c) and (e)] Amplitude response $A(\Omega)$ of the optimal filter designed by the proposed method and
 [(b), (d) and (f)] probability density function $U(\phi)$ of the phase difference $\phi$
 for [(a) and (b)] $q_1(\phi)$, [(c) and (d)] $q_2(\phi)$, and [(e) and (f)] $q_3(\phi)$.
 The insets display snapshots of the oscillators in the $v$--$u$ plane.}
 \label{fig.2}
\end{figure}

In the second experiment, we use the Hodgkin-Huxley (HH) model~\cite{hh}.
It also models periodic firing of a neuron, but it has more realistic, higher-dimensional dynamics
{{without} odd symmetry, in contrast to the FHN model.
We apply the noisy input $I_j(t)$ {as well as} a constant input $I_0=10$ to the 
{$V$} variable (i.e., membrane potential) of the HH model.}
The oscillation frequency is approximately $\omega = 0.438$,
and the sensitivity function and its Fourier coefficients are shown in figs.~\ref{fig.3} (a) and (b).
In addition to $q_1(\phi)$ and $q_2(\phi)$, we use $q_{4}(\phi)$ for optimization with the aim of forming two-clustered states.
%%
%\begin{align}
% q_4(\phi) = \cos2\phi,
 %%
% \quad q_5(\phi) = \frac{1}{2}[\delta(\phi) + \delta(\phi-\pi)]. \label{eq. q2}
%\end{align}
%%{$q_4(\phi) = \cos2\phi.$}
%%
The power spectra $P_{\xi}(\Omega)$, $P_{\eta}(\Omega)$ and $P_{\zeta}(\Omega)$, the  parameters $C$ and $\alpha$, and the basis function $W(\Omega)$ are the same as before.
The noise intensity is $\epsilon=0.1$ and the maximum wavenumber of the filter is $m=4$ (because $|z_{\ell}|$ almost vanishes at $\ell=5$).

Figure~\ref{fig.4} shows the numerical results for the HH model.
Synchronized states are successfully formed for $q_{1}(\phi)$ and $q_{2}(\phi)$ as shown in figs.~\ref{fig.4} (b) and (d).
When we use the delta-shaped $q_2(\phi)$, we obtain a nontrivial filter consisting of multiple modes [figs.~\ref{fig.4} (c)]
and the PDF $U(\phi)$ has a sharper peak than the case with the sinusoidal $q_1(\phi)$ [figs.~\ref{fig.4} (d)].
In contrast to the FHN model, we can realize {a} two-clustered state as shown in figs.~\ref{fig.4} (f),
because $Z(\theta)$ of the HH model has a sufficiently large second Fourier component as shown in fig.~\ref{fig.3} (b).
{Note that the realizability of a particular state is determined by the sensitivity function $Z(\theta)$ that characterizes the phase response property of {the driven} oscillator,
rather than by the dimensionality or complexity of the oscillator model.}

\begin{figure}
\onefigure[width=8cm]{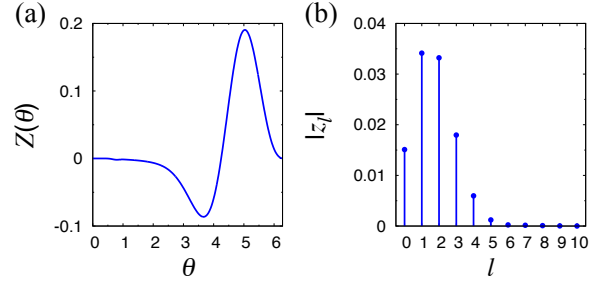}
 \caption{Hodgkin-Huxley model.
 (a) Sensitivity function $Z(\theta)$ and
 (b) Fourier coefficients.}
 \label{fig.3}
\end{figure}

\begin{figure}
\onefigure[width=8cm]{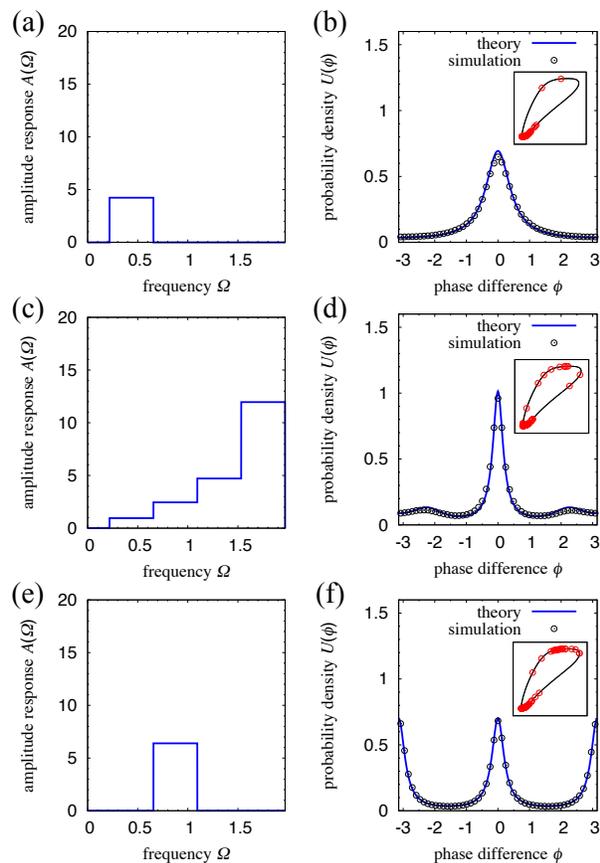}
 \caption{Hodgkin-Huxley model.
 [(a), (c) and (e)] Amplitude response $A(\Omega)$ of the optimal filter designed by the proposed method and
 [(b), (d) and (f)] probability density function $U(\phi)$ of the phase difference $\phi$
 for [(a) and (b)] $q_1(\phi)$, [(c) and (d)] $q_2(\phi)$ and [(e) and (f)] $q_4(\phi)$.
 The insets display snapshots of the oscillators in the $V$--$m$ plane, where $m$ is a channel variable~\cite{hh}.}
 \label{fig.4}
\end{figure}

\section{Summary and discussion}

We have proposed a method for designing and controlling various noise-induced synchronization patterns by filtering the input noise,
including the synchronized and clustered states.
By numerical simulations, the validity of the method has been confirmed for two types of limit-cycle oscillators.
\INS{These results will provide a theoretical basis for optimizing noise-induced synchronization by filtering the input noise.}

Though some previous works~\cite{abouzeid2009,hata2011} proposed optimization methods for the phase response property of the oscillator to enhance noise-induced synchronization, {those works} considered only the Lyapunov exponent of the phase (i.e., the exponential decay rate of the small phase difference between two oscillators),
so that they could not fully characterize the synchronized states and could not be used to design various synchronization patterns as described in this letter.
More importantly, in contrast to previous works~\cite{abouzeid2009,hata2011} that gave the optimal phase response property of the oscillator,
our present study provides a method to generate optimal noisy inputs to the oscillator, which can be implemented much more easily than designing the oscillator response. 
\INS{Thus, our method can be useful in various real-world applications, e.g.,
energy-efficient synchronization control in wireless sensor networks~\cite{yasuda2013}.}
%%
%%
%%\INS{
%%In contrast to conventional control methods for synchronization patterns by phase locking, e.g., ref.~\cite{kori2008}, our method will be useful in real-world applications utilizing common environmental noise for synchronization as in some plants~\cite{koenig1998}. For example,
%%In ref.~\cite{yasuda2013}, common environmental noise is utilized for energy-efficient synchronization control in wireless sensor networks.}
%For example, our method could be used to artificially design the synchronization patterns of rhythmic elements such as spiking neurons. 
%%
%It has been reported that the synchronization of olfactory bulb neurons is induced by correlated noisy inputs
%from inhibitory presynaptic neurons~\cite{galan06__correl_induc_synch_of_oscil}.
%%
%By injecting artificially generated noisy inputs to the neurons, their synchronization patterns may be controlled.
%%
%Our method could also be useful in engineering applications of the noise-induced synchronization, such as synchronization of wireless sensor networks by environmental signals as considered in ref.~\cite{yasuda2013}.
%%
%%We expect that our method {will} provide general formulation and theoretical basis for such applications.
%%\INS{In a sense, this result might indicate the theoretical possibility of similar filtering mechanisms in biological systems such as~\cite{koenig1998}.}

{Finally, though we have not considered the effect of differences in the natural frequency of the oscillators~\cite{yoshimura08__invar_of_frequen_differ_in,burton2012,zhou2013} in this letter,
it is often significant in practical applications. Extension of the present method to non-identical oscillators will be an important future work.}

\acknowledgments

The authors are grateful to H.~Yasuda,  M.~Harashima, Y.~Honda, Y.~Horio, and K.~Aihara for fruitful discussions. 
Financial support by KAKENHI (25540108, 26103510, 26120513), CREST Kokubu project of JST, and FIRST Aihara project of JSPS are gratefully acknowledged.

\end{document}